\documentclass[preprint,12pt]{article}

\usepackage[usenames]{color}

\usepackage[figuresright]{rotating}

\def\be{\begin{equation}}
\def\ee{\end{equation}}
\def\bea{\begin{eqnarray}}
\def\eea{\end{eqnarray}}

\usepackage{amssymb}
\usepackage{framed}
\usepackage{graphicx}
\usepackage{multirow}
\usepackage{caption}
\usepackage{subfig}
\begin{document}
\title
{In what sense a neutron star$-$black hole binary is the holy grail for testing gravity?}
\author{Manjari Bagchi$^1$ and Diego F. Torres$^2$\\ \\
{\small $^1$International Centre for Theoretical Sciences} \\ 
{\small Tata Institute of Fundamental Research,  Bangalore 560012, India}\\
{\small $^2$ICREA \& Institute of Space Sciences, Barcelona 2a Planta E-08193, Spain }\\
{\small manjari.bagchi@icts.res.in, dtorres@ieec.uab.es}}

\maketitle
\begin{center}
Received Honorable Mention in the 2014 Awards for Essays on Gravitation \\ by the Gravity Research Foundation
\end{center}

\vskip1.0cm
\begin{abstract}

Pulsars  in binary systems have been very successful to test the validity of general relativity in the strong field regime \cite{stairs03, stairs04, kramer06, stairs10}. So far, such binaries include neutron star$-$white dwarf (NS-WD) and neutron star$-$neutron star (NS-NS) 
systems. It is commonly believed that a neutron star$-$black hole (NS-BH) 
binary will be much superior for this purpose. But in what sense is this true? Does it apply to all possible deviations?

\end{abstract}

No NS-BH system is known yet, although different studies on such possible binaries in the Galactic disk \cite{ppr05, kh09}, globular clusters \cite{sig03}, and the Galactic center \cite{fl11} have been presented. If NS-BH binaries have small orbital periods ($P_b \lesssim 1$ day), as predicted \cite{ppr05, kh09}, how better can we test gravity? Is there any aspect for which they would not be the preferred probe? \\

The Parametrized Post-Newtonian (PPN) formalism is very useful to denote differences among theories of gravity via the values of a few parameters.\footnote{These are denoted by $\gamma^{\rm PPN}$, $\beta^{\rm PPN}$, $\xi$, $\alpha_1^{\rm PPN}$, $\alpha_2^{\rm PPN}$, $\alpha_3^{\rm PPN}$, $\zeta_1$, $\zeta_2$, $\zeta_3$, and $\zeta_4$ at the first order, and $\epsilon$ and $\zeta$ at the second order.} 
These parameters are used to derive expressions for various gravitational effects, including the deflection and delay of light,  the precession of the periastron, and others. 
In the framework of general relativity, $\gamma^{\rm PPN} = \beta^{\rm PPN} = 1$ and all other parameters are zero \cite{will01}. \\

In the strong field regime (as is the case for many binary pulsars), it is conventional to describe the orbital dynamics in terms of five Keplerian and eight post-Keplerian parameters \cite{dd86, kopei94}. The Keplerian parameters are the orbital period $P_b$, the orbital eccentricity $e$, longitude or periastron $\omega$, projected semi-major axis of the orbit $x=a_p \sin i$ ($i$ is the inclination angle between the orbital plane and the sky plane), and the epoch of periastron passage $T_0$. The five most significant post-Keplerian parameters are: 

\begin{enumerate}
\item the advance of the periastron ($\dot{\omega}$), 
\item the decay of the orbital period ($\dot{P_b}$) due to the spin-2 quadrupolar gravitational wave emission ($\dot{P_b}^{\rm Q, spin2}$), proper motion of the pulsar ($\dot{P_b}^{\rm sh}$, known as Shklovskii effect \cite{shl70}), and the acceleration due to the Galactic potential ($\dot{P_b}^{\rm Gal}$) \cite{paulo12},
\item the Shapiro range parameter ($r$), 
\item the Shapiro shape parameter ($s$), 
\item and the Einstein parameter ($\gamma$).
\end{enumerate}

Pulsar astronomers \textit{use the leading order expressions under general relativity for the above mentioned parameters,} as can be found in \cite{lkbook}. Measurements of these post-Keplerian parameters lead to the determination of the masses of the pulsar and the companion\footnote{$\dot{P_b}^{\rm sh}$ and $\dot{P_b}^{\rm Gal}$ are classical effects and should be taken out (whenever significant) from the observed $\dot{P_b}$ before using the general relativistic expression of $\dot{P_b}^{\rm Q, spin2}$.}. Agreement between mass measurements using any three or more of these parameters point to the correctness of the underlying theory. Moreover, the measurement of $\dot{P_b}^{\rm Q, spin2}$ for the case of PSR B1913+16 is considered to be the first indirect detection of gravitational waves \cite{tfm79}. The other post-Keplerian parameters are $\delta_{r}$, $\delta_{\theta}$ (manifestation of the relativistic deformation of the orbit, contributing to the relativistic R\"omer delay) and the aberration parameters $A$, $B$. The aberration parameters can be absorbed into redefinations of $T_0$, $x$, $e$, $\delta_{r}$, and $\delta_{\theta}$ \cite{lkbook}. It is impossible to measure $\delta_{r}$, as it can be absorbed in a redefinition of the rotational phase of the pulsar (page 284 of \cite{dd86}, page 227 of \cite{lkbook}).  $\delta_{\theta}$ is `in principle' possible to measure, but the knowledge of the orientation of the spin axis of the pulsar with respect to the orbital angular momentum is needed (Eqns. 8.67 and 8.70 of \cite{lkbook}). Sometimes, it is possible to determine two other post-Keplerian parameters $\dot{e}$ and $\dot{x}$ \cite{lkbook, damour07}. Emission of gravitational waves results in the decrease in $x$, $P_b$, $e$ \cite{pet64}. Another gravitational effect is the geodetic precession ($\Omega_{geod}$), which is the precession of the spin axes of the pulsar and the companion around the total angular momentum (as the orbital angular momentum is much larger than the spin momenta, it is usually considered as fixed and aligned with the total angular momentum) caused by the curvature of space-time near gravitating bodies (i.e. the pulsar and the companion) \cite{lkbook, boc79}. This geodetic precession affects both $\dot{e}$ and $\dot{x}$ \cite{lkbook}. In addition to gravitational effects, proper motion of the pulsar also contributes to $\dot{\omega}$ and $\dot{x}$ \cite{kopei96}, as it does to $\dot{P_b}$ via $\dot{P_b}^{\rm sh}$ \cite{shl70}.\\

Among the alternative theories of gravity, a prominent place is occupied by the scalar-tensor ones, which assume that in addition of the metric, there are long range scalar fields through which gravity mediates. 
The \textit{coupling strength} between the scalar field $\varphi$ and matter is denoted by (Eqn. 1.3 of  \cite{damespo96}):
\be 
\alpha (\varphi)= \frac{\partial \ln A(\varphi)}{\partial \varphi}, 
\label{eq:alpha_def}
\ee
where $A(\varphi)$ is the `coupling function'. Higher derivatives are denoted by 
$\beta (\varphi) = \frac{\partial \alpha(\varphi)}{\partial \varphi}$, $\beta^{\prime} (\varphi) = \frac{\partial \beta(\varphi)}{\partial \varphi}$, and $\alpha_0 = \alpha (\varphi_0)$, $\beta_0 = \beta (\varphi_0)$, and $\beta^{\prime}_0 = \beta^{\prime} (\varphi_0)$, where $\varphi_0 = \varphi(r \rightarrow \infty)$ is the value at infinity (in the weak field limit) or the cosmological background without the presence of any gravitating body. 
Following \cite{damour07}, we write 
\bea
\gamma^{\rm PPN} = 1 - 2 \frac{\alpha_0^2}{1+\alpha_0^2} \nonumber \\
\beta^{\rm PPN} = 1 + \frac{1}{2} \frac{\alpha_0^2 \beta_0}{(1+\alpha_0^2)^2}. 
\eea
Another useful relation is $\alpha_0^2 = \frac{1}{2 \omega_{\rm BD} +3}$ leading to $\gamma^{\rm PPN} = \frac{\omega_{\rm BD} + 1}{\omega_{\rm BD} + 2}$, where $\omega_{\rm BD}$ is the Brans-Dicke parameter. Measurements of $\gamma^{\rm PPN}$ and $\beta^{\rm PPN}$ constrain $\alpha_0$, $\beta_0$, and $\omega_{\rm BD}$. \\

In case of a binary pulsar, the coupling strengths of the pulsar (mass $M_p$) and the companion (mass $M_c$) can be defined as (page 26 of \cite{damour07}):
\bea 
\alpha_p ({\varphi_{a}}_p) &=& \frac{\partial \ln M_p}{\partial  {\varphi_{a}}_p}, \nonumber \\
\beta_p ({\varphi_{a}}_p) &=& \frac{\partial \alpha_p ({\varphi_{a}}_p)}{\partial {\varphi_{a}}_p}, \nonumber \\
\alpha_c ({\varphi_{a}}_c) &=& \frac{\partial \ln M_c}{\partial  {\varphi_{a}}_c}, \nonumber \\
\beta_c ({\varphi_{a}}_c) &=& \frac{\partial \alpha_c ({\varphi_{a}}_c)}{\partial {\varphi_{a}}_c}.
\label{eq:alphabetadefs}
\eea
${\varphi_{a}}_p$ is the value of $\varphi$ at a large distance from the pulsar, and a combination of $\varphi_{0}$ and the scalar influence of the companion; while ${\varphi_{a}}_c$ is the value of $\varphi$ at a large distance from the companion, and a combination of $\varphi_{0}$ and the scalar influence of the pulsar. Following the convention (page 28 of \cite{damour07}), we define $\alpha_p = \alpha_p (\varphi_0)$, $\alpha_c = \alpha_c (\varphi_0)$, $\beta_p = \beta_p (\varphi_0)$, $\beta_c = \beta_c (\varphi_0)$. $\alpha_c = 0$ for a black hole (`no-scalar-hair' theorem). \\

As the effect of gravity is larger for short orbital periods (except for the Einstein parameter $\gamma$ which scales as $P_b^{1/3}$ \cite{dd86, lkbook}, we select some short period NS-NS and NS-WD  binaries and compare them with a hypothetical NS-BH binary in Table \ref{tab:compareBinaries}. We use these systems to compute leading deviations and testing power. \\

General relativity and many other alternative theories of gravity predict conservation of total angular momentum. A non-zero value of $\alpha_3^{\rm PPN}$ and/or $\zeta_2$ in the PPN formalism implies violation of conservation of total momentum, and is supposed to produce a self acceleration of the center of mass of a binary as \cite{will01}: 
\begin{equation}
{\rm a}_{\rm CM} = \frac{2 \pi^2}{G} (\zeta_2+\alpha_3^{\rm PPN}) \frac{M_p M_c \vert M_p - M_c \vert}{(M_p + M_c)^2} \frac{e}{P_b^2(1-e^2)^{3/2}} ~~.
\end{equation}
In turn, a non-zero value of ${\rm a}_{\rm CM}$ implies a non-zero second time derivative of the orbital period 
$P_b$, for which PSR B1913+16 places a limit $(\zeta_2+\alpha_3^{\rm PPN}) < 4 \times 10^{-5}$ (see page 54 of \cite{will01}).
As the mass factor in the expression of ${\rm a}_{\rm CM}$ is much larger for a NS-BH binary in comparison with a NS-NS star binary, a non-zero value of $(\zeta_2+\alpha_3^{\rm PPN})$ will give a larger value of ${\rm a}_{\rm CM}$ for the case of a NS-BH binary. We compare the contribution from binary parameters in ${\rm a}_{\rm CM}$ in Table \ref{tab:comparePPN}. The larger the contribution from binary parameters, the larger is the value of ${\rm a}_{\rm CM}$ for the same non-zero value of $(\zeta_2+\alpha_3^{\rm PPN})$. It is clear from Table \ref{tab:comparePPN} that the total contribution from the binary parameters (multiplication of the second and the third columns) is much larger for a NS-BH binary in comparison to NS-WD or NS-NS systems. In particular, the contribution from the orbital parameters (third column) is much larger for a NS-NS system than NS-WD systems because of the large eccentricity for the first case. On the other hand, the contribution from the mass factor is larger for a NS-BH binary than a NS-NS binary even if the orbital contribution can be slightly smaller. In summary, as the combined contribution of the orbital and mass parameters in the expression of ${\rm a}_{\rm CM}$ is much larger for a NS-BH binary, such a system will be a very good tool to test the validity of non-conservative theories of gravity. \\

\begin{table}[t]
\caption{\small Parameters for some selected binary pulsars. $M_p$ stands for the pulsar mass, $M_c$ for the companion mass, $P_b$ is the orbital period, and $e$ is the eccentricity.}
\scriptsize
\vspace{-.5cm}
\begin{center}
{\begin{tabular}{lccccc} \hline \hline
 & $M_p$ & $M_c$ & $P_b$ & $e$ &  comments  \\ 
  & (${\rm M_{\odot}}$) & (${\rm M_{\odot}}$) & (day) &  &    \\ \hline
J0348+0432   & 2.01  & 0.17 & 0.10  & $2.36 \times 10^{-6}$  & NS-WD  \\ 
        &   &  &  &  &  Have been used to test theories of gravity \cite{anton12}   \\ \hline
J1738+0333   & 1.46  & 0.18 & 0.35 & $3.4 \times 10^{-7}$  &  NS-WD   \\
        &   &  &  &  & Have been used to test theories of gravity \cite{paulo12}  \\ \hline
J0737-3039A  & 1.34  & 1.25 & 0.10 & 0.09 &  NS-NS    \\
        &   &  &  &  & binary (double pulsar). Provided stringent test     \\ 
        &   &  &  &  & of general relativity in the strong field regime \cite{kramer06}    \\ \hline
NS-BH        & 1.4  & 10.0 & 0.13 & 0.1 & Hypothetical NS-BH   \\
        &   &  &  &  & Agrees with the most probable parameters of \cite{ppr05}.  \\ 
\hline \hline
\end{tabular}}
\end{center}
\label{tab:compareBinaries}
\end{table}

\begin{table}[t]
\caption{\small The contribution from binary parameters in ${\rm a}_{\rm CM}$.}
\scriptsize
\vspace{-.5cm}
\begin{center}
{\begin{tabular}{lcc} \hline \hline
 &  $\frac{M_p M_c \vert M_p - M_c \vert}{(M_p + M_c)^2} $ &  $ \frac{e}{P_b^2(1-e^2)^{3/2}}$ \\ 
  &   &   \\ 
  & (${\rm M_{\odot}}$)  & (${\rm day^{-2}}$)  \\ \hline
J0348+0432  (NS-WD) & 0.13 & $2.36 \times 10^{-4}$ \\ \hline
J1738+0333  (NS-WD) & 0.12 & $2.78 \times 10^{-6}$  \\\hline
J0737-3039A (NS-NS) & 0.02 & 9.11 \\\hline
NS-BH        & 0.93  & 6.01 \\ 
\hline \hline
\end{tabular}}
\end{center}
\label{tab:comparePPN}
\end{table}

\begin{table}[t]
\caption{\small Values of post-Keplerian parameters (measured and predicted) and geodetic precession (predicted). Predicted values of $r$ for the hypothetical NS-BH system is $49.27$ ms ($r=G M_{c}/c^3,~M_c$ being the mass of the companion, $G$ is the gravitational constant and $c$ is the speed of light). Measured values of $r$ for J0348+0432 is $0.85 $ ms, for J1738+0333 is $ 0.89$ ms and for J0737-3039A is $6.21$ ms. Other measured values of the parameters in the table have been marked with $^*$.}
\scriptsize
\vspace{-.5cm}
\begin{center}
{\begin{tabular}{lccccc} \hline \hline
 & $\dot{\omega}$ & $\gamma$  & $\dot{P}_b^{\rm Q, spin2}$ & $\delta_{\theta}$ &$\Omega_{geod}$ \\  
 &   &  &     &  & \\ 
 & (deg/yr)  & (ms) &   ($ 10^{-12} \, {\rm s \ s^{-1}}$)   &  & (deg/yr) \\ \hline
 J0348+0432   & 14.93  & $1.85 \times 10^{-6}$ & $0.259$   & $1.32 \times 10^{-5}$ & 0.77 \\ 
 J1738+0333 & 1.56  & $2.84 \times 10^{-7}$   & $0.027$ & $4.74 \times 10^{-6}$ & 0.11 \\
J0737-3039A  & 16.899$^*$  &$ 0.386^*$ &   $-1.25^*$ & $1.26 \times 10^{-5}$ & 4.78 \\
NS-BH & 30.52  & 2.93   & $-4.34$ & $2.23 \times 10^{-5}$ & 13.93 \\ \hline \hline
 &  $(M_p+M_c)^{2/3}$ & $\frac{M_c(M_p+2 M_c)}{(M_p+M_c)^{4/3}}$   &  $\frac{M_p M_c}{(M_p+M_c)^{1/3}}$ & $\frac{\frac{7}{2} M_p^2 +6 M_p M_c +2 M_c^2}{(M_p+M_c)^{4/3}}$ &$\frac{M_c (4 M_p + 3 M_c)}{(M_p + M_c)^{4/3}}$ \\ 
   &   &  &     &  & \\ 
 &  (${\rm M_{\odot}^{2/3}}$) & (${\rm M_{\odot}^{2/3}}$)  & (${\rm M_{\odot}^{5/3}}$) & (${\rm M_{\odot}^{2/3}}$)  & (${\rm M_{\odot}^{2/3}}$) \\ \hline 
J0348+0432   & 1.68 &  0.14  & 0.27  & 5.75 & 0.52 \\ 
J1738+0333   & 1.39  & 0.17  & 0.22 & 4.71 & 0.60 \\
J0737-3039A  & 1.88 & 1.35  & 1.22 & 5.47 & 3.20 \\
NS-BH & 5.06  & 8.34  & 6.22 & 11.34 & 13.87 \\ 
\hline \hline
\end{tabular}}
\end{center}
\label{tab:comparePKparams}
\end{table}

As already mentioned, so far all measurements of post-Keplerian parameters for binary pulsars considered only the leading order terms under general relativity, and no discrepancy has been seen \cite{stairs03,stairs04,kramer06,stairs10}. For a NS-BH binary, the values of these post-Keplerian parameters will be larger, e.g. the Shapiro range parameter $r$ for a NS-BH binary is more than 7 times larger than that for a NS-NS binary. The deflection of the pulsar beam will also be larger as it scales as $\frac{M_c}{{(M_p+M_c)^{1/3}} \, P_b^{2/3}}$ \cite{dk95} and will be possible to measure. As an example, this factor is 4 times larger for a NS-BH binary than PSR J0737-3039 (Table \ref{tab:compareBinaries}). In the upper panel of Table \ref{tab:comparePKparams}, we compare values of a few post-Keplerian parameters, as well as that of the geodetic precession for different binaries. In the lower panel of the same Table we compare only the mass factors in each term, 
to understand the effect of this factor.\footnote{We exclude $s$ which depends only on the inclination of the orbit with respect to the sky plane.} \textit{It is evident that all the post-Keplerian parameters are larger for NS-BH binaries.} The ratio of the value of $\dot{e}$ (due to the spin-2 quadrupolar gravitational wave emission) for a NS-BH binary to that of a NS-NS binary is around 3. There will be an additional term due to the geodetic precession \cite{lkbook}, which is again around 2 times larger for a NS-BH binary in comparison to a NS-NS binary (provided spin of the neutron stars are the same). Remember, there will be additional contribution from the proper motion of the pulsar in $\dot{\omega}$ \cite{kopei96}.\\

High values of these leading order post-Keplerian parameters for NS-BH systems implies that these terms will be measurable even with a shorter data span. 
Moreover, even the higher order terms might be significant. 
It has been already shown that for such systems, the second post-Newtonian order and the spin-orbit coupling terms will be significant for $\dot{\omega}$ \cite{bagchi13}. 
As one determines masses of the pulsar and the companion by solving the expressions of post-Keplerian parameters with their measured values \cite{lkbook}, neglecting higher orders terms in those expressions (when significant) will lead to inaccurate estimates of masses.
It has also been shown that for a moderately wide range of spin parameters of the black hole and the neutron star, the spin-orbit contribution from the neutron star can be neglected in comparison to that of the black hole \cite{bagchi13}. In such cases, it will be possible to estimate the dimensionless spin parameter of the black hole from this effect, provided the masses of the neutron star and the black hole can be estimated from other post-Keplerian parameters and the spin-orbit contribution is larger than the measurement error in $\dot{\omega}$. Although this spin-orbit coupling will also lead to a very high value of the geodetic precession of the spin axis of the pulsar (Table \ref{tab:comparePKparams}), the precession of the spin axis of the black hole will be much smaller (the mass factor is only 2.42). If the dimensionless spin parameter of the black hole turns out to be larger than unity, it will falsify Penrose's `Cosmic Censorship Conjecture'. Additionally, determination of the quadrupole moment (from the classical spin-orbit coupling) of a black hole in addition to its spin and mass can in principle verify the no-hair theorem \cite{wexkopeikin99}. But measurement of the quadrupole moment is possible only for a massive ($M_{bh} \geq 30 \, {\rm M_{\odot}}$) black hole with short orbital period ($P_b \leq 0.2$ day) binary with a millisecond pulsar \cite{wexliu12}. Similarly, the frame-dragging propagation effect can be measurable with 1 $\mu$s timing accuracy only for short orbital period binaries with orbital inclination very close to 90$^{\circ}$ and a very rapidly rotating black hole (see Eqn. 22 of \cite{wexkopeikin99}). The significance of the higher order terms in the expressions of other post-Keplerian parameters will be worth exploring. \\

For scalar-tensor theories of gravity, in addition to spin-2 quadrupolar gravitational wave emission; spin-0 monopolar, dipolar and quadrupolar gravitational wave emission and the change in the value of the gravitational constant are also responsible for the decay of the orbital period \cite{damour07}.\footnote{Note that additional classical effects in the change of the orbital period, e.g., the Galactic acceleration effect and Shklovskii effect, depend on the location and proper motion of the binary, and can be modeled provided accurate knowledge of these parameters are available \cite{bb96, paulo12}. Commonly, pulsar distances are estimated using the values of their dispersion measures and are limited by the accuracy of the Galactic electron density model \cite{ne2001}; sometimes, distances can be estimated better using paralaxes or HI absorption. Usually, uncertainities in distance estimates place limitations on the testing power of theories of gravity using the observed values of $\dot P_b$ (page 18 of \cite{wex14}). On the other hand, measurements of Shklovskii effect can in principle lead to even better distance estimates when the theory of gravity is known or its effect is negligible \cite{bb96}.} \\

The decay of the orbital period due to the emission of the spin-0 scalar dipolar gravitational waves can be expressed as (Eqn. 86 of \cite{damour07}):

\begin{equation}
\dot{P}_{b}^{\rm D, spin0} = - 4 \pi^2 \frac{G^{*}}{c^3} \frac{1}{P_b} \frac{M_p M_c}{(M_p + M_c)} \frac{1+e^2/2}{(1-e^2)^{5/2}} (\alpha_p -\alpha_c)^2 + \mathcal{O}(1/c^5) + \mathcal{O}(1/c^7)
\label{eq:PbD0}
\end{equation} where $\alpha_0$ is already defined as $\alpha (\varphi_0)$ (after Eqn. \ref{eq:alpha_def}) with $\varphi_0 = \varphi(r \rightarrow \infty)$ is the value of the scalar field $\varphi$ without the presence of any gravitating body. $\alpha_p = \alpha_p (\varphi_0)$, $\alpha_c = \alpha_c (\varphi_0)$, (as defined after Eqn. \ref {eq:alphabetadefs}). For a NS-WD system $\alpha_c = \alpha_{wd} \rightarrow \alpha_0$. For a NS-BH system $\alpha_c = \alpha_{bh} = 0$ (`no-scalar-hair' theorem for black holes). $G^{*} = G/(1+\alpha_0^2)$ is the `bare' gravitational constant, appearing in the action while $G$ is the gravitational constant measured by Cavendish experiment \cite{damour07}. For a NS-NS binary $\alpha_p \simeq \alpha_c$ as the masses, radii and the constituent matter of the pulsar and the companion are almost the same, so we expect their coupling to $\varphi_0$ to be almost the same. So, for the case of a NS-NS binary, $\dot{P}_{b}^{\rm D, spin0}$ is infinitesimally small. For a NS-WD system it can be large if $\alpha_p \gg \alpha_c$ ($\alpha_c = \alpha_{wd} \rightarrow \alpha_0$). The same value of $\alpha_p$ will lead to a larger value of $\dot{P}_{b}^{\rm D, spin0}$ for a NS-BH binary where $\alpha_c = \alpha_{bh} = 0$ (`no-scalar-hair' theorem). Note that, the left hand side of Eqn. \ref{eq:PbD0} contains three unknown parameters, $\alpha_0$, $\alpha_p$ and $\alpha_c$; which changes to two unknowns for a NS-BH system as $\alpha_c = \alpha_{bh} = 0$. Remember, all these coupling terms vanish in general relativity.
As $\dot{P}_{b}^{\rm D, spin0}$ scales as $\mathcal{O}(1/c^3)$ while the spin-0 monopolar and spin-0 quadrupolar emission scale as $\mathcal{O}(1/c^5)$ (as will be clear in Eqns. \ref{eq:Pbdotothers} below), usually these latter terms have been neglected for NS-WD binaries (e.g. \cite{lange01}). \textit{Here we explore whether it is wise to do so for NS-BH binaries.} The expressions for $\dot{P}_{b}^{\rm M, spin0}$, $\dot{P}_{b}^{\rm Q, spin0}$, $\dot{P}_{b}^{\rm Q, spin2}$ with  $\alpha_c = 0$ are as follow \cite{damespo92}:

\bea
\label{eq:Pbdotothers}
\dot{P}_{b}^{\rm M, spin0} &=& - \frac{\pi}{3} \frac{M_p M_c (3 M_p + 5 M_c)^2}{(M_p + M_c)^{7/3}} \left(\frac{G^{*}}{c^3} \right)^{5/3} \left(\frac{2 \pi}{P_b} \right)^{5/3} \frac{e^2 (1+e^2/4)}{(1-e^2)^{7/2}} \alpha_p^2 + \mathcal{O}(1/c^7) \\
\dot{P}_{b}^{\rm Q, spin0} &=& - \frac{32 \pi}{5} \frac{M_p M_c^3}{(M_p+ M_c)^{7/3}} \left(\frac{G^{*}}{c^3} \right)^{5/3} \left(\frac{2 \pi}{P_b} \right)^{5/3} \frac{(1+73e^2/24 + 37 e^4/96)}{(1-e^2)^{7/2}} \alpha_p^2 + \mathcal{O}(1/c^7)  \nonumber \\
\dot{P}_{b}^{\rm Q, spin2} &=& - \frac{192 \pi}{5} \frac{M_p M_c}{(M_p+ M_c)^{1/3}} \left(\frac{G^{*}}{c^3} \right)^{5/3} \left(\frac{2 \pi}{P_b} \right)^{5/3} \frac{(1+73e^2/24 + 37 e^4/96)}{(1-e^2)^{7/2}} + \mathcal{O}(1/c^7) \nonumber 
\eea
Note that here $\dot{P}_{b}^{\rm Q, spin2}$ takes the familiar form as in general relativity with $G$ replaced by $G^{*}$ (not the case if $\alpha_c \neq 0$). In Table \ref{tab:comparePdots} we compare different terms for a NS-BH binary for different values of $\alpha_0$, i.e. for different values of $G^*$. It is clear that although $\dot{P}_{b}^{\rm D, spin0}$ is few orders of magnitude larger than other terms, those should not be neglected, in particular if one wants to constrain values of $\alpha_0$ and $\alpha_{p}$. The excess orbital decay (after taking out $\dot{P}_{b}^{\rm Q, spin2}$ and the Galactic acceleration and the Shklovskii terms) should be modelled as the sum of $\dot{P}_{b}^{\rm M, spin0}$, $\dot{P}_{b}^{\rm D, spin0}$ and $\dot{P}_{b}^{\rm Q, spin0}$. Now, to perform an order of magnitude comparison of the effect of different types of gravitational wave emission from different types of binaries, in Table \ref{tab:comparePdots2} we assume $\alpha_{wd} = \alpha_{0} = 0$. It is clear that unlike NS-BH systems, the contributions from the monopolar and quadrupolar (both spin-0 and spin-2) emissions are much smaller in comparison to the dipolar emission for NS-WD systems, thus making NS-WD systems better tools to detect the dipolar gravitational wave, if it exists, i.e. general relativity is violated. \\

\begin{table}[t]
\caption{\small The rate of decay of orbital period of binary pulsars due to different types of gravitational waves emissions, for a standard NS-BH binary. We display many significant digits only to demonstrate the effect of different values of $\alpha_0$.}
\scriptsize
\vspace{-.5cm}
\begin{center}
{\begin{tabular}{l|cc|cc|cc|cc} \hline \hline
  & \multicolumn{2}{|c|}{$\dot{P}_{b}^{\rm M, spin0} $}    & \multicolumn{2}{|c|}{$\dot{P}_{b}^{\rm D, spin0}$}  & \multicolumn{2}{|c|}{$\dot{P}_{b}^{\rm Q, spin0}$} & \multicolumn{2}{|c}{$\dot{P}_{b}^{\rm Q, spin2}$} \\ 
    & \multicolumn{2}{|c|}{($ 10^{-15} \, \alpha_{p}^{-2} \, {\rm s s^{-1}}$)}  & \multicolumn{2}{|c|}{($ 10^{-8} \, \alpha_{p}^{-2} \, {\rm s s^{-1}}$) } & \multicolumn{2}{|c|}{($10^{-13} \, \alpha_{p}^{-2} \, {\rm s s^{-1}}$)} & \multicolumn{2}{|c}{($ 10^{-12} \, {\rm s s^{-1}}$)} \\ 
   & $\alpha_0 = 10^{-2}$ & $\alpha_0 = 10^{-4}$ & $\alpha_0 = 10^{-2}$ & $\alpha_0 = 10^{-4}$ & $\alpha_0 = 10^{-2}$& $\alpha_0 = 10^{-4}$& $\alpha_0 = 10^{-2}$ & $\alpha_0 = 10^{-4}$ \\ \hline
NS-BH        & $ -8.28237  $  & $ -8.28375 $ & $ -2.19145 $ & $ -2.19166 $ & $ - 5.56419  $ & $ -5.56511 $ & $ -4.33873 $ & $ 4.33945 $ \\ 
\hline \hline
\end{tabular}}
\end{center}
\label{tab:comparePdots}
\end{table}

\begin{table}[b]
\caption{\small The rate of decay of orbital period of binary pulsars due to different types of gravitational waves emissions, for different types of binaries.}
\scriptsize
\vspace{-.5cm}
\begin{center}
{\begin{tabular}{lcccc} \hline \hline
  & $\dot{P}_{b}^{\rm M, spin0} $  & $\dot{P}_{b}^{\rm D, spin0}$  & $\dot{P}_{b}^{\rm Q, spin0}$ & $\dot{P}_{b}^{\rm Q, spin2}$ \\ 
       &   &  &  & \\ 
   & ($\alpha_{p}^{-2} \, {\rm s s^{-1}}$)  & ($\alpha_{p}^{-2} \, {\rm s s^{-1}}$)  & ($\alpha_{p}^{-2} \, {\rm s s^{-1}}$) & ($ {\rm s s^{-1}}$) \\ \hline
J0348+0432  (NS-WD) & $-1.28 \times 10^{-25}$  & $ - 3.53 \times 10^{-9}$ & $-2.70 \times 10^{-16}$ & $-2.59 \times 10^{-13}$ \\ \hline
J1738+0333  (NS-WD) & $-2.90 \times 10^{-28}$ & $- 1.03 \times 10^{-9}$ & $-5.61 \times 10^{-17}$ & $-2.72 \times 10^{-14}$ \\\hline
NS-BH        & $ -8.28 \times 10^{-15}$  & $ - 2.19 \times 10^{-8}$ & $-5.56 \times 10^{-13}$ & $-4.34 \times 10^{-12}$\\ 
\hline \hline
\end{tabular}}
\end{center}
\label{tab:comparePdots2}
\end{table}

Similarly, extra terms involving $\alpha_p$ and $\alpha_c$ appear in the leading order expression of other post-Keplerian parameters like $r, \,s, \, \gamma, \, \dot{\omega}$ (page 29 and 30 of \cite{damour07})\footnote{Remember the effect of proper motion in $\dot{\omega}$ which is a classical effect and remains the same irrespective of the validity of general relativity \cite{kopei96}.}. These expressions also become similar to those under general relativity with $G$ replaced by $G^{*}$ for a NS-BH binary. \\

Another fascinating potential of use of binary pulsars is the possibility of detection of the variation of $G$ with time. The rate of change of orbital period due to the change in the value of $G$ (if any), can be expressed as \cite{paulo12}:
\begin{equation}
\dot{P}_{b}^{\rm \dot{G}} = -2 \frac{\dot{G}}{G} \left[ 1 - \frac{2q+3}{2q+2}\varsigma_p - \frac{3q+2}{2q+2}\varsigma_c \right] P_b
\label{eq:GGdot}
\end{equation} where $q=M_p/M_c$ is the mass ratio, and $\varsigma = \frac{\partial ( \ln M)}{\partial ( \ln G)} = \frac{E_{grav}}{M c^2}$ is the `sensitivity' or the `compactness'. $\varsigma_p$ is the compactness of the pulsar and $\varsigma_c$ is the compactness of the companion. So, for a NS-NS system $\varsigma_p = \varsigma_c = \varsigma_{ns}$, for a NS-WD system $\varsigma_p = \varsigma_{ns}$, $\varsigma_c = \varsigma_{wd}$, and for a NS-BH system $\varsigma_p = \varsigma_{ns}$, $\varsigma_c = \varsigma_{bh}$. Usually, $\varsigma_{ns} \simeq 0.15$ (depends somewhat on the equation of the state), $\varsigma_{bh} = 0.5$, $\varsigma_{wd} \simeq 10^{-4}$. In Table \ref{tab:compareGGdot}, we compare the terms inside the square bracket in Eqn. (\ref{eq:GGdot}), from which it is clear that NS-BH systems are not good to detect any change in the value of $G$.\\

\begin{table}[t]
\caption{\small The contribution from mass and compactness in the expression of $\dot{P}_{b}^{\rm \dot{G}}$, using $\varsigma_{ns} \simeq 0.15$, $\varsigma_{bh} = 0.5$, $\varsigma_{wd} \simeq 10^{-4}$.}
\scriptsize
\vspace{-.5cm}
\begin{center}
{\begin{tabular}{l|c} \hline \hline
 &  $ 1 - \frac{2q+3}{2q+2}\varsigma_p - \frac{3q+2}{2q+2}\varsigma_c $  \\  \hline
 J0348+0432  (NS-WD) & 0.844  \\ \hline
J1738+0333  (NS-WD) &  0.842  \\\hline
J0737-3039A (NS-NS) &  0.625 \\\hline
NS-BH               &  0.254  \\ 
\hline \hline
\end{tabular}}
\end{center}
\label{tab:compareGGdot}
\end{table}

We now explore the possibility of verification of the strong equivalence principle with different types of binary pulsars. The parameter relevant for the strong equivalence principle is $\Delta_i$ for the object `$i$' \cite{stairs03}:
\begin{equation}
\left( \frac{M_{grav}}{M_{intertial}} \right)_i = 1 + \Delta_i = 1 + \eta \varsigma_i + \eta^{\prime} \varsigma_i^2 + \ldots
\label{eq:sep}
\end{equation} 
where 
\bea
\eta = 4 \beta^{\rm PPN} - \gamma^{\rm PPN} - 3 - 10 \xi/3 - \alpha_1 + 2 \alpha_2/3 -2 \zeta_1/3 - \zeta_2/3, 
\label{eq:etadef}
\eea
$\eta^{\prime}$ is a combination of 1PPN and 2PPN parameters \cite{damour07}, and $\varsigma_i$ is the compactness of the object `$i$' as defined after Eqn. (\ref{eq:GGdot}). For a binary pulsar, $\Delta_{net} = \vert \Delta_p - \Delta_c \vert$ will give an acceleration to the center of mass of the binary. It is obvious that $\Delta_{net}$ will be much larger for NS-WD binaries in comparison to NS-NS or even NS-BH binaries because of the large difference in the values of $\varsigma_{ns}$ and $\varsigma_{wd}$, making NS-WD binaries the best tools to verify the strong equivalence principle. Some of the practical methods to verify the strong equivalence principle have been discussed in section 4 of \cite{wex14}. \\

In summary, higher order post-Newtonian terms will be essential to incorporate while modelling orbital dynamics of NS-BH binaries (Table \ref{tab:comparePKparams} and related discussion). These systems might help to determine the spin parameter of the BH and test the validity of `Cosmic Censorship Conjecture'. As these are also important sources for gravitational waves for Advanced LIGO, understanding of the properties of these binaries from pulsar data analysis will help the gravitational wave community to build better waveform templates. In this article, we have demonstrated that these systems will be better tools to test the validity of non-conservative theories of gravity which predict a self acceleration of the center of mass of the binary (Table \ref{tab:comparePPN}). From Table \ref{tab:comparePdots2} we see that NS-BH systems will be better systems to detect deviations from general relativity, as the combined effect of spin-0 monopolar, dipolar and  qudrupolar gravitational wave emission is much larger than that of NS-WD systems. But one should be more careful while modelling such systems by considering all these effects, whereas for the case of NS-WD systems, only dipolar emission dominates (if exist at all) and the other effects can be neglected (as done by \cite{lange01}). We have also seen that NS-BH systems will not be as good as NS-WD systems to test alternative theories of gravity by detecting the change in the value of the gravitational constant (Table \ref{tab:compareGGdot}), or a violation of the strong equivalence principle (discussion after Eqn. \ref{eq:etadef}). \\

Finally, we should remember that the presence of red noise (which appears for a long data span) makes measurements of Keplerian and post-Keplerian parameters difficult and erroneous \cite{koppot00}. Use of `Cholesky' method is helpful in such situation \cite{chc11} and the present day pulsar timing analysis package `tempo2' is equipped with this method\footnote{http://www.atnf.csiro.au/research/pulsar/tempo2/index.php?n=Main.ModelSpectraDetail}   \\

{\small The authors thank the anonymous referee for useful comments on the manuscript. DFT acknowledges support from grant AYA 2012-39303 and SGR2014-1073.}



\end{document}